\documentclass[12pt,preprint]{aastex}
\usepackage{amssymb}

\usepackage{amsmath}

\usepackage{graphics}

\newcommand{\be}{\begin{equation}}
\newcommand{\ee}{\end{equation}}
\newcommand{\bd}{\begin{displaymath}}
\newcommand{\ed}{\end{displaymath}}

\slugcomment{submitted to ApJ, \today}

\shorttitle{The relation of optical/UV and X-ray emission in LLAGN }
\shortauthors{Y-D Xu }

\begin{document}

\title{The relation of optical/UV and X-ray emission in low-luminosity active galactic nuclei }

\author{Ya-Di Xu}
\affil{Physics Department, Shanghai Jiao Tong University, 800
Dongchuan Road, Shanghai 200240, China\\
Email: ydxu@sjtu.edu.cn}
 \clearpage

\begin{abstract}

We study the relation of optical/UV and X-ray emission in the low
luminosity active galactic nuclei (LLAGNs), using a sample of 49
sources including 28 local Seyfert galaxies and 21 low-ionization
nuclear emission-line regions (LINERs) with the optical/UV spectral
luminosity at the wavelength $\lambda=2500~ {\rm \AA}$, $23.0\leq
\log L_{\nu({\rm 2500}~{\rm \AA})}~({\rm erg~s^{-1}}~{\rm
Hz^{-1}})\leq 27.7$, and the X-ray spectral luminosity at 2 keV,
$20.5\leq \log L_{\nu(2~{\rm keV})}\leq 25.3$. The strong
correlations are found between the X-ray luminosity and the
optical/UV to X-ray index, $\alpha_{\rm ox}$, with the optical/UV
luminosity, with the slopes very similar to the findings for the
luminous AGNs in the previous works. The correlation between
$\alpha_{\rm ox}$ and $L_{\nu(2~{\rm keV})}$ is very weak as that
found for the luminous AGNs in the majority of previous similar
works. We also study the relation between $\alpha_{\rm ox}$ and the
Eddington ratio $L_{\rm bol}/L_{\rm Edd}$ for our sample and find a
significant anti-correlation for the sources with $L_{\rm
bol}/L_{\rm Edd}\lesssim 10^{-3}$, which is opposite to the
correlation between the two variables for the luminous AGNs. Using
the advection dominated accretion flow (ADAF) model, we roughly
reproduce this anti-correlationship for the two variables for the
LLAGNs. This result strongly supports the ADAF as a candidate
accretion mode in LLAGNs.

\end{abstract}

\section{Introduction}

In the long term and various types of studies on the active galactic
nuclei (AGNs), the spectral energy distribution (SED) is one of the
most focused topics investigated in both the observational and
theoretical works, which is believed to provide clues to the
physical mechanism of the emission from AGNs. There are many
different types of active galaxies, such as Seyfert galaxies, radio
galaxies, quasars, and low ionized nuclear emission-line region
(LINER) galaxies, etc., which have different emission properties.
Although the detailed physical mechanism is undetermined, it is
commonly accepted that they probably have very different accretion
modes which result in their different emission spectra. The
optical/UV continuum in the luminous quasar is supposed to be the
blackbody radiation from the thin accretion disk surrounding the
black hole in AGN, while the X-ray emission is originated from the
hot corona located above the thin disk where the disk seed photons
are inverse Compton up-scattered by the energetic electrons
\citep*[e.g.,][]{1999MNRAS.303L..11Z, 2002ApJ...572...79C,
2002MNRAS.331L..35F, 2007MNRAS.381.1426M, 2009MNRAS.394..207C,
2010arXiv1012.0439V}. For low-luminosity AGNs (LLAGNs), namely
LINERs and local Seyfert galaxies, it is still unclear whether the
accretion mode is the same as luminous AGNs or a different accretion
mode and physical mechanism is present. By comparing the nuclear
spectral energy distribution (SED) of 13 nearby LINERs with the
average SED of powerful quasars, \citet{2007MNRAS.377.1696M} found
that the broad band SEDs of LINERs are quite similar to the SED of
more luminous AGN. \citet{2008A&A...490..995P} studied the near-IR
to X-ray spectrum of four low luminosity Seyfert 1 galaxies and
concluded these LLAGNs have the same shape as the spectrum of
quasars that are $10^2-10^5$ times more luminous, which suggested
that the thin accretion disk plus hot corona model may still sustain
at low accretion rate in the LLAGNs. However,
\citet{1999ApJ...516..672H} found that the LLAGN SEDs have a weak or
absent blue bump which is a typical emission characterizing the
blackbody radiation from the thin accretion disk.
Advection-dominated accretion flows (ADAFs) have been suggested to
be present in many black hole systems accreting at low rates,
including some quiescent X-ray binaries and LLAGNs, to reproduce the
spectral energy distributions (SEDs) from these sources\citep*
[e.g.,][]{ny94,ny95,lackny96,nmy96,nmq98}. \citet{qdnh99} showed
that the optical/UV to X-ray emission detected from the nuclei of
M81 and NGC 4579 can be well explained by an optically thick,
geometrically thin accretion disk which extends down to $\sim 100
R_{\rm S}$ ($R_{\rm S}=2GM/c^2$, inside which an ADAF is present)
\citep{2009RAA.....9..401X}. \citet{2004ApJ...606..173P} found that
ADAF models can be used to fit the UV to X-ray SED of a LINER
galaxy, NGC 3998 quite well. Moreover, \citet{2009MNRAS.399..349G}
found a significant anticorrelation between the hard X-ray photon
index $\Gamma$ and the Eddington ratio $\L_{\rm bol}/L_{\rm Edd}$
for a sample of LLAGNs \citep*[see also][]{2011A&A...530A.149Y},
which is in contrast with the positive correlation for luminous AGNs
\citep*{{2004ApJ...607L.107W},2006ApJ...646L..29S,2008ApJ...682...81S},
but similar to that of X-ray binaries (XRBs) in the low/hard state
\citep*{2005ChJAS...5..273Y,{2007ApJ...658..282Y}}. They suggested
that the accretion mode in LLAGNs may be similar to that of XRBs in
the low/hard state, of which the X-ray emission is assumed to
originate from the Comptonization process in ADAF.

The relations between the quantities characterizing the optical/UV
and X-ray properties of the observed AGN SEDs have been
comprehensively studied. Some studies suggested that the dependence
of the optical/UV to X-ray spectral indices between the rest-frame
2500 ${\rm \AA}$ and 2 keV, $\alpha_{\rm ox}$, on optical/UV
luminosity may be primarily caused by the dependence on redshift
\citep*{2003ApJ...588..119B,2007ApJ...657..116K}, and some others
argued that this dependence may be artificially resulted from the
larger dispersion of the optical luminosities deviating from the
average SED than that of the X-ray luminosities
\citep*{1998A&A...334..498Y,{2007MNRAS.377.1113T}}. Many recent
works found the significant relations between $\alpha_{\rm ox}$ (and
X-ray luminosity) with optical/UV luminosity in different wide
luminosity and redshift range AGN samples
\citep*[e.g.,][]{2003AJ....125..433V,2005AJ....130..387S,2006AJ....131.2826S,
{2007ApJ...665.1004J},{2010ApJ...708.1388Y},2010ApJS..187...64G,
2010A&A...512A..34L}. Most of these previous works focused on the
luminous AGN sources, with the optical luminosity $L_{\nu (2500 {\rm
\AA})}\gtrsim 10^{27}~{\rm erg~s}^{-1}~{\rm Hz^{-1}}$ and the
Eddington luminosity ratio $L_{\rm bol}/L_{\rm Edd}\gtrsim 10^{-3}$,
which showed that $\alpha_{\rm ox}$ is positively correlated with
the Eddington luminosity ratio for the different samples. Following
the spectral evolution of a galactic black hole binary, GRO
J1655-40, \citet{2011MNRAS.413.2259S} simulated the spectral states
of AGN and modeled SEDs for a mixture of AGNs in different spectral
states, which predicted that the correlations between $\alpha_{\rm
ox}$ and the Eddington luminosity ratio changes the sign when the
the Eddington luminosity ratio changes from above to below a
critical value,  $L_{\rm bol}/L_{\rm Edd}\sim 10^{-2}$.

In this work, we use a sample of 49 LLAGNs to explore the relations
between $\alpha_{\rm ox}$ and X-ray luminosity with optical/UV
luminosity, and the relation between $\alpha_{\rm ox}$ and the
Eddington luminosity ratio $L_{\rm bol}/L_{\rm Edd}$ in LLAGNs, and
compare the results with those of previous works. In addition, we
try to model the SEDs of LLAGNs in different accretion rates and
different black hole masses with ADAF models, and compare the
theoretical results with the observational ones.  The sample and the
estimates of the optical luminosity for the sources are described in
\S 2. In \S 3, we present the results for the sample. We use the
ADAF model to explain the statistic results of the LLAGNs in \S 4
and the discussion is in \S 5. The cosmological parameters
$H_{0}=70~\rm km~s^{-1} Mpc^{-1}$, $\Omega_{M} = 0.3$ and
$\Omega_{\Lambda} = 0.7$ have been adopted in this work.

\section{The Sample}

For the purpose of our study on LLAGN, we construct a sample of 49
AGN sources including 21 LINERs and 28 local Seyfert galaxies, which
originate from a sample compiled by \citet{2009MNRAS.399..349G}. The
objects in \citet{2009MNRAS.399..349G}'s sample are drawn from the
Palomar sample \citep{1997ApJS..112..315H,{2008ARA&A..46..475H}} and
the multi-wavelength catalogue of LINER (MCL)
\citep{1999RMxAA..35..187C}. To complete their study on the relation
between the hard X-ray photon index and the Eddington ratio in
LLAGN, they carefully selected sources for the sample. The X-ray
observation with Chandra and XMM-Newton were searched for the
objects. The sources without the nuclei at hard X-ray band ($>2$
keV) were excluded to ensure that the X-ray emission are really from
the nuclear component of the AGN. The Compton-thick sources were
also excluded because the Compton-thick absorption may flat the
X-ray photon index in 2-10 keV spectral fitting which leads error in
the study. In addition, the measurement of black hole mass or
stellar velocity dispersion were also one of the selected criteria
for the sample. They finally compiled a sample of 55 sources, of
which 27 are LINERs and 28 are local Seyfert galaxies. These 28
local Seyfert galaxies are totally selected for our sample. For the
27 LINERS, we select 21 of them for our sample, the remaining 6
LINERs are excluded because the observational data used to calculate
the optical/UV luminosity for the study of this work are
unavailable.

For the selected sources in our sample, the redshift $z$, the black
hole mass $M_{\rm BH}$, the spectral photon index $\Gamma$, and the
integrated 2-10 keV luminosity for every sources are all available
in \citet{2009MNRAS.399..349G} for our present investigation. To
obtain the optical/UV luminosity, we use several different
approaches for the sources which have different observational data.
For 19 Seyferts, we use the observed absolute B magnitude of the
nucleus \citep{2001ApJ...555..650H} to compute the optical/UV
luminosity of the source combined with the observed distance
(labeled as approach 1). Analyzing the high-resolution images of the
objects observed by the Hubble Space Telescopes (HST),
\citet{2001ApJ...555..650H} obtained the nuclear magnitudes and
luminosities of H$\beta$ lines including narrow plus broad
components. The absolute magnitudes of the nuclei have been
corrected for Galactic extinction but not for internal extinction.
\citet{2001ApJ...555..650H} also presented a relation between the
optical continuum and Balmer emission line luminosity which is valid
for both luminous and low-luminosity AGNs.

For the other 9 Seyferts and 3 LINERs, we use the data of optical
emission lines, H$\alpha$ (narrow plus broad component) luminosity
\citep{1997ApJS..112..391H} and the ratio of H$\alpha$ to H$\beta$
\citep{1997ApJS..112..315H}, to compute the luminosity of $H\beta$
(labeled as approach 2). For other 7 LINERs, only narrow component
of H$\alpha$ is available \citep{1997ApJS..112..315H}, but the ratio
of broad H$\alpha$ to total (broad plus narrow component) H$\alpha$
\citep{1997ApJS..112..391H} and the ratio of H$\alpha$ to H$\beta$
\citep{1997ApJS..112..315H} are also available, thus we can
calculate the total H$\beta$ luminosity (labeled as approach 3). For
the left 11 LINERs, we have narrow H$\alpha$ luminosity and the
ratio of H$\alpha$ to H$\beta$ for every source
\citep{1997ApJS..112..315H}, but the ratio of broad H$\alpha$ to
total H$\alpha$ is not available. The averaged value from the former
7 LINERs are used (labeled as approach 4), which should be
reasonable. The absolute B magnitudes for the above sources can then
be obtained according to the relation of $L_{\rm H\beta}$ and
$M_{B}$ \citep{2001ApJ...555..650H}
\begin{equation}
\log L_{\rm H\beta}=(-0.34\pm 0.012)M_{B}+(35.1\pm 0.25).
\end{equation}
With the corrected absolute B magnitudes and the distances derived
from the redshift, we can then calculate the B band spectral fluxes
and luminosities from the nuclei of the sources for this work. A
typical spectral index for the featureless optical/UV continuum of
the Seyfert 1 nuclei \citep*[e.g.,][]{1987ApJ...315...74W},
$\alpha_{\rm o}=-1$, is adopted in the calculation. All the data for
our sample are listed in Table 1.

\section{Calculation of the optical/UV-to-X-ray indices for the sample }

When the B band optical luminosity is derived from the absolute B
magnitude, we can extrapolate the optical continuum to the UV with
the typical optical spectral index, $\alpha_{\rm o}=-1$, and obtain
the spectral luminosity at the wavelength $\lambda=2500 {\rm \AA}$,
$L_{\nu(2500 {\rm \AA})}$. The X-ray spectral luminosity at 2 keV ,
$L_{\nu(2 {\rm keV})}$, can be computed with the known integrated
2-10 keV luminosity and the hard X-ray photon index compiled from
\citet{2009MNRAS.399..349G}. Figure 1 displays the relations between
the the X-ray spectral luminosity  $L_{\nu(2 {\rm keV})}$ and the
optical/UV spectral luminosity at the wavelength $\lambda=2500 {\rm
\AA}$, $L_{\nu(2500 {\rm \AA})}$, of our sample. We find a
significant correlation for the two quantities as follows:
\begin{equation}
\log L_{\nu(2 {\rm keV})}=(0.652\pm 0.082)\log L_{\nu(2500 {\rm
\AA})}+(6.269\pm 2.044),
\end{equation}
with the probability of a null coefficient $P<10^{-6}$. The
optical/UV and X-ray spectral luminosities of our sample ranged in
$\sim 10^{23.0-27.7}~{\rm erg~s}^{-1}$ and $\sim 10^{20.5-25.3}~{\rm
erg~s}^{-1}$, respectively.

The optical/UV-to-X-ray spectral index, which is traditionally
defined as
\begin{equation}
\alpha_{\rm ox}=-\frac {\log L_{\nu(2500 {\rm \AA})}/L_{\nu(2{\rm
keV})}}{\log \nu(2500 {\rm \AA})/\nu(2 \rm{keV})}, \label{def_alpha}
\end{equation}
can be calculated with optical/UV and X-ray spectral luminosities.
We list the derived optical/UV luminosity, X-ray spectral
luminosity, and $\alpha_{\rm ox}$ of our sample in Table 1.

The relation between $\alpha_{\rm ox}$ and $L_{\nu(2500 {\rm \AA})}$
for the sources in our sample is also studied as many previous
similar works. The result is shown in Figure 2. The
optical/UV-to-X-ray spectral index, $\alpha_{\rm ox}$, lies from
$\sim 0.2$ to $1.6$ for the most sources. We find a significant
correlation between $\alpha_{\rm ox}$ and $\log L_{\nu(2500 {\rm
\AA})}$, the significance level for disapproving the null hypothesis
that the two variables are uncorrelated is less than $7\times
10^{-4}$ for all the sources. The best-fitting linear regression
line for the sample is
\begin{equation}
\alpha_{\rm ox}=(0.134\pm 0.031)\log L_{\nu(2500 \rm \AA)}-(2.406\pm
0.785),
\end{equation}
which is also plotted in Figure 2. The fitted $\alpha_{\rm ox}-\log
L_{\nu(2500 \rm \AA)}$ slope is consistent with that inferred from
the relation of $\log L_{\nu(2 {\rm keV})}-\log L_{\nu(2500 \rm
\AA)}$ and the definition of $\alpha_{\rm ox}$ in equation
\ref{def_alpha}.

We also investigate the relationship between $\alpha_{\rm ox}$ and
the X-ray spectral luminosity at 2 keV, $L_{\nu(2 {\rm keV})}$. The
significance level for approving that the two variables are
correlated is about $67.7\%$ for $\alpha_{\rm ox}-\log L_{\nu(2 {\rm
keV})}$. Moreover, the relationship between $\alpha_{\rm ox}$ and
the Eddington ratio $L_{\rm bol}/L_{\rm Edd}$ is explored, where the
bolometric luminosity is calculated from the integrated 2-10 keV
X-ray luminosity assuming that $L_{\rm bol}/L_{(2-10 {\rm keV})}\sim
30$ as done by \citet{2009MNRAS.399..349G}. For all the sources of
our sample, we find that the significance level for approving that
the two variables are anti-correlated is about $61.6\%$ for
$\alpha_{\rm ox}-\log L_{\rm bol}/L_{\rm Edd}$. {According to the
ADAF model, the optically thin ADAF solution exists only for the
mass accretion rate $\dot{m}$ (defined as
$\dot{m}=\dot{M}/\dot{M_{\rm Edd}}$, $\dot{M_{\rm Edd}}=L_{\rm
Edd}/0.1c^{2}$) less than a critical value $\dot{m}_{\rm crit}$,
which is a dependence of the viscosity parameter $\alpha$,
$\dot{m}_{\rm crit}\sim
0.3\alpha^{2}$\citep{{1997ApJ...477..585M},{nmq98}}. For the adopted
parameter $\alpha=0.2$ in our calculations (see next section for the
details), the corresponding value of the critical mass accretion
rate is $\dot{m}_{\rm crit}\sim 0.01$, above which no global ADAF
solution is available. From our theoretical calculations with ADAF
model (see next section for the details), we find that the ADAF
accreting at $\dot{m}_{\rm crit}\sim 0.01$ corresponds to the
Eddingtong ratio $L_{\rm bol}/L_{\rm Edd}\sim 10^{-3}$. Thus,} we
re-test the relation of $\alpha_{\rm ox}-\log L_{\rm bol}/L_{\rm
Edd}$ for the sources with $L_{\rm bol}/L_{\rm Edd}\lesssim
10^{-3}$. After excluding the sources with the Eddingtong ratio
$L_{\rm bol}/L_{\rm Edd}> 10^{-3}$, the significance level for
approving that the two variables are anti-correlated is improved to
about $97.3\%$. The best-fitting linear regression line for 36
sources with the Eddingtong ratio $L_{\rm bol}/L_{\rm Edd}\lesssim
10^{-3}$ is
\begin{equation}
\alpha_{\rm ox}=(-0.163\pm 0.070)\log L_{\rm bol}/L_{\rm
Edd}+(0.185\pm 0.321).
\end{equation}
Figures 3 and 4 show the relations of $\alpha_{\rm ox}-\log L_{\nu(2
{\rm keV})}$ and $\alpha_{\rm ox}-\log L_{\rm bol}/L_{\rm Edd}$,
respectively.

\section{The optical/UV-to-X-ray indices simulated with the ADAF model}

The ADAFs are supposed to be present in LLAGNs, of which the
accretion rate is very low, e.g., below the critical value,
\textbf{$\dot{m}_{\rm crit}\sim 0.3\alpha^{2}$}. The global
structure of an ADAF surrounding a black hole with mass $M_{\rm bh}$
can be calculated, if some parameters, $\dot{m}$, $\alpha$, $\beta$
and $\delta$, are specified. We employ the approach suggested by
\citet{m00} to calculate the global structure of an accretion flow
surrounding a Schwarzschild black hole in the general relativistic
frame. All the radiation processes are included in the calculations
of the global accretion flow structure \citep*[see][for details and
the references therein]{m00,{2009ApJ...699..722Y}}. The values of
parameters adopted in this work are different from those in
\citet{m00}. The value of the viscosity parameter $\alpha$ is still
a controversial issue \citep*[see][for details and the references
therein]{2011ApJ...729...10X}. In this work, a typical value of the
viscosity parameter $\alpha= 0.2$ is adopted in the calculations.
The parameter $\beta$, defined as $p_{\rm
m}=B^2/8\pi=(1-\beta)p_{\rm tot}$, ($p_{\rm tot}=p_{\rm gas}+p_{\rm
m}$), describes the magnetic field strength of the accretion flow.
We assume $\beta= 0.8$ in all the calculations. This parameter will
mainly affect the radio spectrum from the source, while it affects
little on the optical/UV and X-ray emission, which we mostly focus
on in this work. The parameter $\delta$ describes the fraction of
the viscously dissipated energy directly going into electrons in the
accretion flow. It was pointed out that a significant fraction of
the viscously dissipated energy could go into electrons by magnetic
reconnection, if the magnetic fields in the flow are strong
\citep{bl97,bl00}. The value of $\delta$ is still uncertain, usually
between 0.1-0.5 \citep{cao05,cao07}, and we adopt a conventional
value of $\delta=0.3$ in all the calculations.

Given the values for parameters $\alpha$, $\beta$ and $\delta$, the
mass accretion rate $\dot{m}$ is changed from $10^{-5}$ to $10^{-2}$
to simulate the LLAGNs with different Eddington ratios. The typical
black hole masses $10^7$, $10^8$, and $10^9$ are adopted for the
LLAGNs in the calculations. The global structure of the ADAF is
obtained, with which the spectrum  of LLAGNs are calculated.

From the derived spectrum of every simulated LLAGN, the optical/UV
to X-ray indices $\alpha_{\rm ox}$ and the Eddington luminosity
ratio $L_{\rm bol}/L_{\rm Edd}$ can then be calculated. To compare
the simulated results with the sample, the bolometric luminosity of
the simulated AGN is also calculated from the integrated 2-10 keV
X-ray luminosity assuming that $L_{\rm bol}/L_{(2-10 {\rm keV})}\sim
30$. The relations of $\alpha_{\rm ox}$ with $\log L_{\rm
bol}/L_{\rm Edd}$ are plotted in different types of lines in Figure
4 for different black hole masses $m_{\rm BH}=10^7, 10^8,$ and
$10^9$.

\section{Discussion}

In this paper, we have explored the relations between $\alpha_{\rm
ox}$ and the X-ray luminosity with the optical/UV luminosity, and
the relation between $\alpha_{\rm ox}$ and the Eddington luminosity
ratio $L_{\rm bol}/L_{\rm Edd}$ for a sample of 49 LLAGNs, including
21 LINERs and 28 local Seyfert galaxies.

The observed significant correlations in $\log L_{\nu(2 {\rm
keV})}-\log L_{\nu(2500 \rm \AA)}$ and $\alpha_{\rm ox} -\log
L_{\nu(2500 \rm \AA)}$ for our LLAGN sample are similar to those
results given in the previous works for luminous AGNs
\citep*[e.g.,][]{2003AJ....125..433V,2005AJ....130..387S,2006AJ....131.2826S,
{2007ApJ...665.1004J},{2010ApJ...708.1388Y},2010ApJS..187...64G,
2010A&A...512A..34L}. The slope of $\log L_{\nu(2 {\rm keV})}-\log
L_{\nu(2500 \rm \AA)}$ relation we obtained in this work is 0.652,
an intermediate value between the slopes, $\sim 0.645-0.760$
\citep{2005AJ....130..387S,2010A&A...512A..34L}, presented in the
previous works. The slope of $\alpha_{\rm ox} -\log L_{\nu(2500 \rm
\AA)}$ relation we obtained in this work is 0.134, an intermediate
value between the slopes, $\sim 0.114-0.154$
\citep{2010ApJS..187...64G,2010A&A...512A..34L}, derived in the
previous works. These correlations between $\log L_{\nu(2 {\rm
keV})}-\log L_{\nu(2500 \rm \AA)}$ and $\alpha_{\rm ox} -\log
L_{\nu(2500 \rm \AA)}$ imply that the more luminous AGN sources emit
more both in the optical/UV and X-ray bands, while the X-ray
emission increases more quickly than the optical/UV emission, if we
believed that these correlations are intrinsic in the SEDs of AGNs
rather than the artificial results caused by the different
dispersions in the optical/UV and X-ray luminosities for the sample.
Although the thin disk plus hot corona model is commonly accepted to
be the possible accretion model for the luminous AGNs to
quantitatively explain the wide band emission,  of which the
optical/UV continuum is supposed to be the blackbody radiation from
the thin disk while the X-ray emission originates from the hot
corona where the disk seed photons are inverse Compton up-scattered
by the energetic electrons, more quantitatively theoretical
calculations about the optical/UV and X-ray emission correlations
are needed to support these arguments. On the other hand, the
similarity of the correlation  in $\log L_{\nu(2 {\rm keV})}-\log
L_{\nu(2500 \rm \AA)}$ and $\alpha_{\rm ox} -\log L_{\nu(2500 \rm
\AA)}$ for our LLAGN sample comparing with the luminous AGN samples
in previous works can not rule out the possibility that there are
different accretion modes presented in the LLAGNs from that in the
luminous AGNs, if the accretion mode in the LLAGNs can also satisfy
these correlations. The ADAF model suggested to be present in the
low luminosity accretion systems is one of the choices. According to
the theoretical calculations as stated in \S 4, we find that the
simulated LLAGN SEDs with the ADAF model can also qualitatively
explain the correlations in $\log L_{\nu(2 {\rm keV})}-\log
L_{\nu(2500 \rm \AA)}$ and $\alpha_{\rm ox} -\log L_{\nu(2500 \rm
\AA)}$ we derived for our LLAGN sample (see Figure
\ref{fig_nu_lnu}). In the ADAF model, the optical/UV emission is
mainly contributed from the inverse Compton scatter of soft
synchrotron photons by the hot electron in the ADAF, and X-ray
emission is contributed both from the inverse Compton scattering and
the bremsstrahlung emission. For the luminous sources with high
accretion rate, the inverse Compton component dominates the X-ray
spectrum. More quantitative theoretical works based on the
distribution functions for the AGNs as functions of various
quantities such as black hole mass, accretion rate, etc., will
provide more convincing explanations to the observed correlations,
which is worthy to be studied in the future work.

The derived correlation of $\alpha_{\rm ox}-\log L_{\rm bol}/L_{\rm
Edd}$, with the slope of such as 0.397 (Lusso 2010), and 0.11 (Grupe
2010), for the luminous AGNs with $L_{\rm bol}/L_{\rm Edd}>
10^{-3}$-$10^{-2}$ in the previous works changes the sign to be
anti-correlation as \citet{2011MNRAS.413.2259S} found, with the
slope of -0.163,  for the sources with $L_{\rm bol}/L_{\rm
Edd}\lesssim 10^{-3}$ in our LLAGN sample. This change of sign
implies that the accretion mode for the LLAGNs may be different from
the one for luminous AGNs. Thus, we simulate some LLAGNs SEDs with
the ADAF model, and study the relation of $\alpha_{\rm ox}-\log
L_{\rm bol}/L_{\rm Edd}$ for the simulated low luminosity sources.
Changing the mass accretion rate from $10^{-4}$ to $10^{-2}$ for the
sources with a given black hole mass, we find that the simulated
optical/UV to X-ray indices decreases with the increasing mass
accretion rate, predicting a strong anti-correlation between the two
variables as showed in our analysis for the LLAGN sample (see Figure
\ref{fig_alpha_redd}). It is needed to notice that the corresponding
$L_{\rm bol}/L_{\rm Edd}$ value is from $\sim 10^{-7}$ to $\sim
10^{-3}$. According to many previous works \citep*[e.g.,][]{n96,
nmy96, emn97, e98, cao03, 2009RAA.....9..401X}, ADAFs may co-exist
with the standard thin disks in the sources when mass accretion
rates are slightly lower than the critical value, e.g., $\dot{m}\sim
10^{-3}$-$10^{-2}$. In this case, the ADAF is present in the inner
region near the black hole and connects to a standard thin disk
(plus hot corona) at a certain transition radius. The contribution
of the outer thin disk (and hot corona) to the continuum may be
important, which will change the optical/UV to X-ray index of the
spectra. This may be the reason that an anti-correlation between
$\alpha_{\rm ox}$ and $\log L_{\rm bol}/L_{\rm Edd}$ is present for
the sources with $L_{\rm bol}/L_{\rm Edd}\lesssim 10^{-3}$. In
addition, we also vary the black hole mass in the simulation, the
relations are almost unchanged because the Eddington luminosity
ratio is insensitive to the black hole mass. Although the
theoretical relation curve is a bit higher than the fitted line for
the sample, our simulation provide a strong support to the ADAF
model as a candidate accretion mode in the LLAGNs. Our sample
contains a small number of LLAGNs. More accumulated LLAGN sources
will improve the work and strengthen our conclusions.

{There have been claims that jet models can reproduce the SEDs of
LLAGN such as Sgr A* and M81
\citep*[e.g.,][]{{2000A&A...362..113F},{2002A&A...383..854Y},{2008ApJ...681..905M}}.
For the sources having jets, there should be some contribution from
the jets to the X-ray emission, which may change the optical/UV to
X-ray spectral index of these sources studied in this work.
Prediction of the relation between the optical/UV to X-ray index and
the Eddington fraction of LLAGNs with jet models can be compared
with the observational statistical findings, which may give us a
clue to the jet models. However, we still lack of certain evidence
showing the presence of jets in most of LLAGN sources in our sample
from both the theoretical and observational researches. Moreover,
for those LLAGN sources probably possessing jets, it is shown that
the dominance of the X-ray emission from the ADAF or jets depends on
the X-ray luminosity of the
source\citep{2005ApJ...629..408Y,2007ApJ...669...96W,{2011arXiv1104.3235W}}.
When the X-ray luminosity is below the critical value ($\sim
10^{-5}-10^{-6}L_{\rm Edd}$), the X-ray emission from the jet should
be the dominant emission rather than that from the accretion flow.
While the X-ray luminosity of the majority sources of our sample is
higher than $\sim 10^{-6}L_{\rm Edd}$, the X-ray emission from the
ADAF should be the major contribution to the source. On the other
hand, there are some additional free parameters describing the jet
model than the pure ADAF model, e.g., the power normalization
$N_{j}$, temperature of particles entering base of jet $T_{\rm e}$,
energy index of electron power-law tail $p$ \citep*[see][for the
details
]{{1995A&A...293..665F},{2000A&A...362..113F},{2002A&A...383..854Y},{2008ApJ...681..905M}}.
The dependence of the calculated spectrum on the model parameters
and their interdependence are complex, and the values of the
parameters are still uncertain. Prediction of the relation between
the optical/UV to X-ray index and the Eddington fraction of LLAGNs
with the jet model is somewhat dependent of the model parameters,
which is beyond the scope of this work.}

{In this work, the bolometric luminosity is estimated from the
integrated 2-10 keV X-ray luminosity assuming a constant bolometric
correction, $L_{\rm bol}/L_{\rm 2-10 keV}=30$, which is determined
from a mean energy distribution calculated from 47 luminous, mostly
nearby quasars \citep{1994ApJS...95....1E}. However, there is
evidence that the bolometric correction depends on the Eddington
ratio in AGN
\citep*[e.g.,][]{{2007MNRAS.381.1235V},{2009MNRAS.392.1124V},{2010ApJ...708.1388Y}}.
\citet{2007MNRAS.381.1235V} suggested a more well-defined
relationship between the bolometric correction and Eddington ratio
in AGN, with a transitional region at an Eddington ratio of $\sim
0.1$, below which the bolometric correction is typically 15-25, and
above which it is typically 40-70. The constant bolometric
correction, $\kappa_{2-10 {\rm keV}}=30$, we adopted in this work is
typical for luminous AGNs with Eddington ratio $L_{\rm bol}/L_{\rm
Edd}\simeq 0.1$. Since the Eddington ratio (or the X-ray 2-10 keV
luminosity) of the LLAGN sources in our sample covers over five
orders of magnitude and are all below $\sim 0.1$, the bolometric
correction and the calculated bolometric luminosity may be smaller
than what we show in Table 1 and Figure 4, if the dependence of the
bolometric correction on the Eddington ratio in AGN does exist.
Moreover, the lower the Eddington ratio (or the X-ray 2-10 keV
luminosity), the smaller the bolometric correction. Thus, the slope
of the correlation between $\alpha_{\rm ox}$ and $L_{\rm bol}/L_{\rm
Edd}$ may be smaller than that obtained in this work.}

{The optical/UV-to-X-ray index, $\alpha_{\rm ox}$, for most of the
sample spans the range $0.6 \lesssim \alpha_{\rm ox} \lesssim 1.4$,
very similar to the results of \citet{2007MNRAS.377.1696M} ($0.8\sim
1.4$) and \citet{2010ApJS..187..135E} ($0.55\sim 1.36$). However,
the scatter of the sources is obvious and nonnegligible, with 8
objects having $\alpha_{\rm ox} \lesssim 0.6$ and 4 objects having
$\alpha_{\rm ox} \gtrsim 1.4$. It appears that the scatter is not
random (see Figures 2 and 4), the objects with $\alpha_{\rm
ox}\lesssim 0.6$ are mostly Seyferts and with $\alpha_{\rm
ox}\gtrsim 1.4$ are mostly LINERs. The sources of this scatter are
not very certain. There are several possible sources which may
originate from several factors. (1) The scatter of the Seyferts with
very low $\alpha_{\rm ox}\lesssim 0.6$ may be due to the difficulty
of the measurements. Even with HST resolution, it is in practice
very difficult to reliably measure nuclear continumm magnitudes for
very faint sources. In our sample, only lower limit magnitudes are
available for five optically faint Seyferts, e.g., NGC 2639, 4138,
4168, 4258, and 7479, and the optical/UV to X-ray indices of these
faint sources are all less than 0.6. Note that the larger magnitudes
will result in lower optical luminosity and higher optical/UV to
X-ray indices for the faint sources, and increase the scatter in the
figures. (2) The optical/UV spectral luminosity of the LINERs are
all calculated from the Balmer emission-line measurements
(approaches 2, 3, and 4) using the $L_{\rm H\beta}\sim M_{\rm B}$
relation derived by fitting the combined sample including PG quasars
and local Seyfert galaxies\citep{2001ApJ...555..650H}. It is found
that the relation for Seyferts appears marginally shallower than
that for quasars \citep*[see the dashed line in Figure 6
of][]{2001ApJ...555..650H}. Therefore, the absolute B magnitude of
the low luminosity sources may be overestimated with a steeper
$L_{\rm H\beta}\sim M_{\rm B}$ relation, and systematically result
in a larger $\alpha_{\rm ox}$ for the objects calculated with
approaches 2, 3, and 4 including all the LINERs and 9 Seyferts. For
the five faint Seyfert galaxies NGC 2639, 4138, 4168, 4258, and
7479, we also use the $L_{\rm H\beta}\sim M_{\rm B}$ relation to
re-estimate the optical/UV spectral luminosity and the optical/UV to
X-ray indices (approach 2), the calculated $\alpha_{\rm ox}$ becomes
larger and in the range $\sim 0.5-1.5$. Thus, the $\alpha_{\rm ox}$
of 19 Seyferts calculated with approach 1 may be systematically
lower than those of objects calculated with approaches 2, 3, and 4.
Moreover, the total line luminosity of $H\beta$, $L_{H\beta}$,
including broad and narrow components is computed with different
types of line data in approaches 2, 3, and 4 (see section 2 for
details), which may also introduce scatter of $\alpha_{\rm ox}$ in
the result. (3) The X-ray data of Seyfert galaxies are all from the
same study using the homogeneous analysis, while that of LINERs are
from various works with the different spectral analysis
\citep*[see][and reference therein]{{2009MNRAS.399..349G}}. This
makes it very difficult to analyze the systematic scatter of
$\alpha_{\rm ox}$ of these LINERs due to its inhomogeneity.  }

\acknowledgments We thank the anonymous referee for very helpful
comments and constructive suggestions. This work is supported by the
NSFC (grants 11078014).

{}



\clearpage
\begin{deluxetable}{ccccccccc}
\tabletypesize{\scriptsize} \tablecaption{The sample.}
\tablewidth{0pt}
\tablehead{\colhead {Name} & \colhead {$z$ } &\colhead {log $M_{\rm
BH}/M_{\odot}$} &\colhead {$f_{\nu}^{\rm B}$} &\colhead
{$\alpha_{\rm ox}$} &\colhead {$l_{\nu({\rm 2~ keV})}$} &\colhead
{$l_{\nu({\rm 2500 \rm \AA})}$} &\colhead {log $L_{\rm bol}/L_{\rm
Edd}$} &\colhead {approach}
\\
\colhead{(1)} & \colhead{(2)} & \colhead{(3)} & \colhead{(4)} &
\colhead{(5)} & \colhead{(6)} & \colhead{(7)} & \colhead{(8)} &
\colhead{(9)}  } \startdata
&&&&Seyfert galaxies&&&&\\
\hline
NGC 1275 & 0.017559 &   8.51 & -25.92 & 1.06   &  24.92 & 27.68 &-2.30 &1\\
NGC 2639 & 0.011128 &   8.02 &$<$-29.02~~ & $<$0.53~~   &  22.80 & $<$24.18~~ &-3.82 &1\\
NGC 3031 &-0.000113$^{a}$ &   7.80 & -26.03 & 0.98   &  22.32 & 24.89 &-4.17 &1\\
NGC 3227 & 0.003859 &   7.59 & -25.86 & 1.05   &  23.66 & 26.41 &-2.47 &1\\
NGC 3516 & 0.008836 &   7.36 & -25.86 & 1.15   &  24.14 & 27.14 &-1.70 &1\\
NGC 4051 & 0.002336 &   6.11 & -26.11 & 0.98   &  23.16 & 25.72 &-1.42 &1\\
NGC 4138 & 0.002962 &   7.75 & $<$-28.10~~ & $<$0.28~~   &  23.21 & $<$23.94~~ &-3.08 &1\\
NGC 4151 & 0.003319 &   7.18 & -24.58 & 1.25   &  24.31 & 27.56 &-1.33 &1\\
NGC 4168 & 0.007388 &   7.95 & $<$-29.86~~ & $<$0.38~~   &  21.98 & $<$22.97~~ &-4.70 &1\\
NGC 4258 & 0.001494 &   7.61 & $<$-28.08~~ & $<$0.24~~   &  22.75 & $<$23.36~~ &-3.37 &1\\
NGC 4388 & 0.008419 &   6.80 & -26.82 & 0.93   &  23.72 & 26.14 &-1.70 &1\\
NGC 4395 & 0.001064 &   5.04 & -26.99 & 0.96   &  21.65 & 24.16 &-1.85 &1\\
NGC 4565 & 0.004103 &   7.70 & -27.54 & 1.28   &  21.47 & 24.79 &-4.89 &1\\
NGC 4579 & 0.005067 &   7.78 & -26.96 & 0.94   &  23.10 & 25.55 &-3.37 &1\\
NGC 4639 & 0.003395 &   6.85 & -27.97 & 0.74   &  22.26 & 24.19 &-3.25 &1\\
NGC 5033 & 0.002919 &   7.30 & -26.37 & 0.99   &  23.08 & 25.66 &-2.85 &1\\
NGC 5273 & 0.003549 &   6.51 & -26.67 & 0.87   &  23.27 & 25.53 &-1.77 &1\\
NGC 5548 & 0.017175 &   8.03 & -26.41 & 0.74   &  25.25 & 27.16 &-1.40 &1\\
NGC 7479 & 0.007942 &   7.07 & $<$-28.39~~ & $<$0.52~~   &  23.16 & $<$24.51~~ &-2.57 &1\\
NGC 2655 & 0.004670 &   7.77 & -27.43 & 0.52   &  23.65 & 25.01 &-2.54 &2\\
NGC 2685 & 0.002945 &   7.15 & -28.08 & 0.97   &  21.42 & 23.96 &-3.82 &2\\
NGC 3147 & 0.009407 &   8.79 & -27.93 & 0.45   &  23.96 & 25.12 &-3.52 &2\\
NGC 3486 & 0.002272 &   6.14 & -28.75 & 0.97   &  20.53 & 23.06 &-3.89 &2\\
NGC 3941 & 0.003095 &   8.15 & -28.30 & 1.06   &  21.03 & 23.78 &-5.89 &2\\
NGC 4477 & 0.004520 &   7.92 & -28.12 & 0.99   &  21.72 & 24.29 &-4.89 &2\\
NGC 4501 & 0.007609 &   7.90 & -28.07 & 1.26   &  21.51 & 24.79 &-4.92 &2\\
NGC 4698 & 0.003342 &   7.84 & -28.41 & 0.95   &  21.27 & 23.74 &-5.30 &2\\
NGC 4725 & 0.004023 &   7.49 & -28.78 & 0.99   &  20.96 & 23.53 &-5.22 &2\\
\hline
&&&&LINERs&&&&\\
\hline
NGC 266  & 0.015547 &   7.90 & -27.99 & 1.05   &  22.76 & 25.50 &-3.64 &3\\
NGC 0315 & 0.016485 &   9.24 & -27.99 & 0.77   &  23.56 & 25.55 &-4.22 &3\\
NGC 2681 & 0.002308 &   7.20 & -27.73 & 1.21   &  20.95 & 24.10 &-4.88 &3\\
NGC 3226 & 0.003839 &   8.24 & -27.80 & 0.59   &  22.92 & 24.47 &-4.12 &3\\
NGC 3718 & 0.003312 &   7.97 & -28.31 & 0.71   &  21.97 & 23.83 &-4.53 &3\\
NGC 4143 & 0.003196 &   8.31 & -27.44 & 1.02   &  22.03 & 24.67 &-4.89 &3\\
NGC 4278 & 0.002165 &   9.20 & -27.05 & 1.07   &  21.94 & 24.72 &-5.86 &3\\
NGC 3169 & 0.004130 &   7.95 & -27.94 & 0.26   &  23.71 & 24.39 &-3.16 &4\\
NGC 4261 & 0.007465 &   8.94 & -28.04 & 0.80   &  22.73 & 24.81 &-4.41 &4\\
NGC 4374 & 0.003536 &   8.80 & -27.92 & 0.99   &  21.71 & 24.28 &-5.82 &4\\
NGC 4457 & 0.002942 &   7.00 & -27.04 & 1.54   &  20.99 & 25.00 &-4.63 &4\\
NGC 4494 & 0.004483 &   7.60 & -28.53 & 1.09   &  21.04 & 23.88 &-5.22 &4\\
NGC 4548 & 0.001621 &   7.51 & -28.09 & 0.63   &  21.79 & 23.43 &-4.34 &4\\
NGC 4552 & 0.001071 &   8.50 & -28.16 & 0.59   &  21.45 & 23.00 &-5.71 &4\\
NGC 4594 & 0.003639 &   9.04 & -26.95 & 1.27   &  21.96 & 25.27 &-5.59 &4\\
NGC 4736 & 0.001027 &   7.42 & -27.87 & 0.96   &  20.76 & 23.25 &-5.39 &4\\
NGC 5746 & 0.005751 &   7.49 & -29.07 & 0.64   &  21.88 & 23.55 &-4.04 &4\\
NGC 6500 & 0.010017 &   8.28 & -26.82 & 1.58   &  22.18 & 26.29 &-5.16 &4\\
UGC 08696 & 0.037780  &  7.74 & -26.57 & 1.35  &  24.18 & 27.70 &-2.18 &2\\
NGC 6240 & 0.024480 &   9.11 & -26.16 & 1.52   &  23.77 & 27.73 &-3.69 &2\\
NGC 7130 & 0.016151 &   7.54 & -26.30 & 1.71   &  22.76 & 27.23 &-3.67 &2\\
\hline
\enddata
\tablecomments{Col.(1): Source name. Col.(2): Redshift.  Col.(3):
Black hole mass. Col.(4): $f_{\nu}^{\rm B}\equiv \log F_{\nu}^{\rm
B}$, the logarithmic spectral flux at B band in units of ${\rm
erg}~{\rm s}^{-1}~{\rm cm}^{-2}~{\rm Hz}^{-1}$. Col.(5): Derived
optical/UV to X-ray luminosity index. Col. (6): $l_{\nu({\rm 2~
keV})}\equiv \log L_{\nu({\rm 2~ keV})}$, the logarithmic X-ray
spectral luminosity at 2 keV in units of ${\rm erg~s}^{-1}~{\rm
Hz}^{-1}$. Col. (7): $l_{\nu({\rm 2500 \rm \AA})}\equiv \log
L_{\nu({\rm 2500 \rm \AA})}$, the logarithmic optical/UV spectral
luminosity at the wavelength $\lambda=2500~\rm \AA$  in units of
${\rm erg~s}^{-1}~{\rm Hz}^{-1}$. Col.(8): Eddington luminosity
ratio. Col.(9): Approach number. The numbers 1, 2, 3, and 4, refer
to the different approaches we used to obtain the absolute B
magnitude for the sources with different usable data, see the text
in \S 2 for the
details. \\
$^{a}$ Distance of 3.5 Mpc.}

\end{deluxetable}

\begin{figure}
\plotone{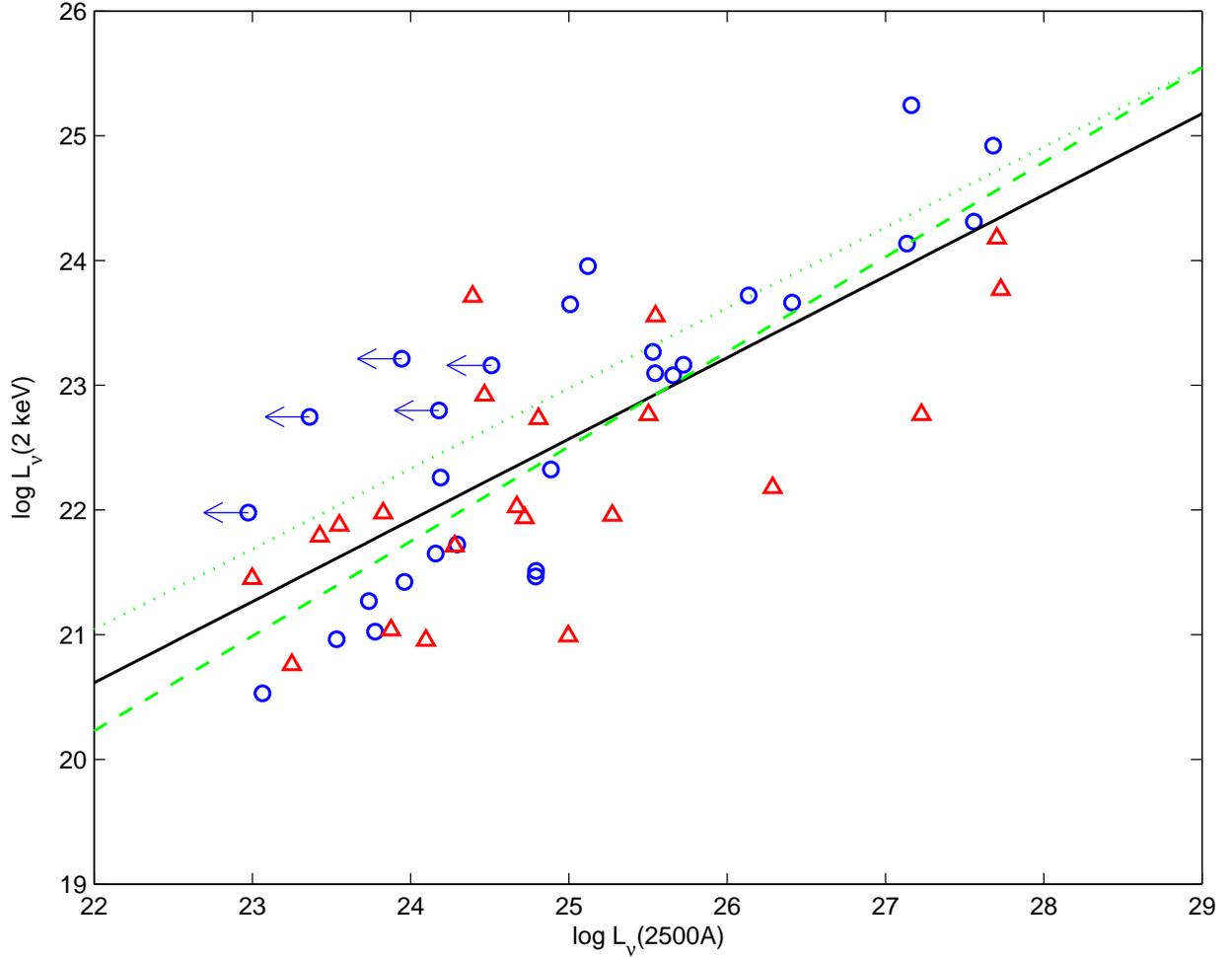} \caption{Optical/UV spectral luminosity at the
wavelength $\lambda=2500~\rm \AA$, $L_{\nu({\rm 2500 \rm \AA})}$,
vs. the X-ray spectral luminosity at 2 keV, $L_{\nu({\rm 2~ keV})}$.
Blue circles and  red triangles refer to the Seyferts  and LINERs,
respectively.  The best fitted line is plotted in solid line. The
green dashed and dotted lines correspond to the correlations found
by Lusso(2010) and Strateva(2005), respectively.}
\label{fig_lx_lopt}
\end{figure}

\begin{figure}
\plotone{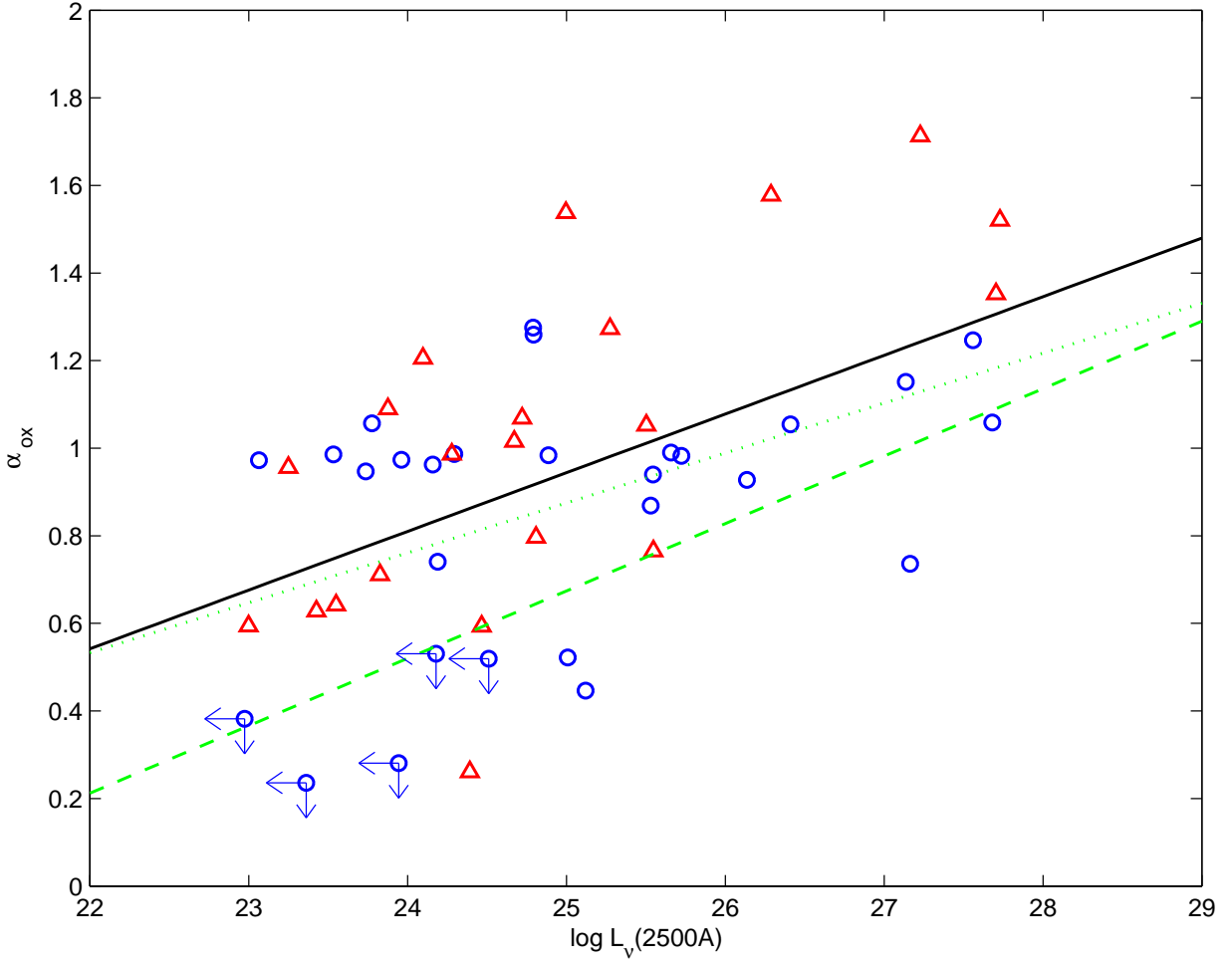} \caption{Optical/UV to X-ray spectral index
$\alpha_{\rm ox}$ vs. the optical/UV spectral luminosity at the
wavelength $\lambda=2500~\rm \AA$, $L_{\nu({\rm 2500 \rm \AA})}$.
The symbols are the same as Figure 1. The best fitted line is
plotted in solid line. The green dashed and dotted lines correspond
to the correlations found by Lusso(2010) and Grupe(2010),
respectively.} \label{fig_alpha_lopt}
\end{figure}

\begin{figure}
\plotone{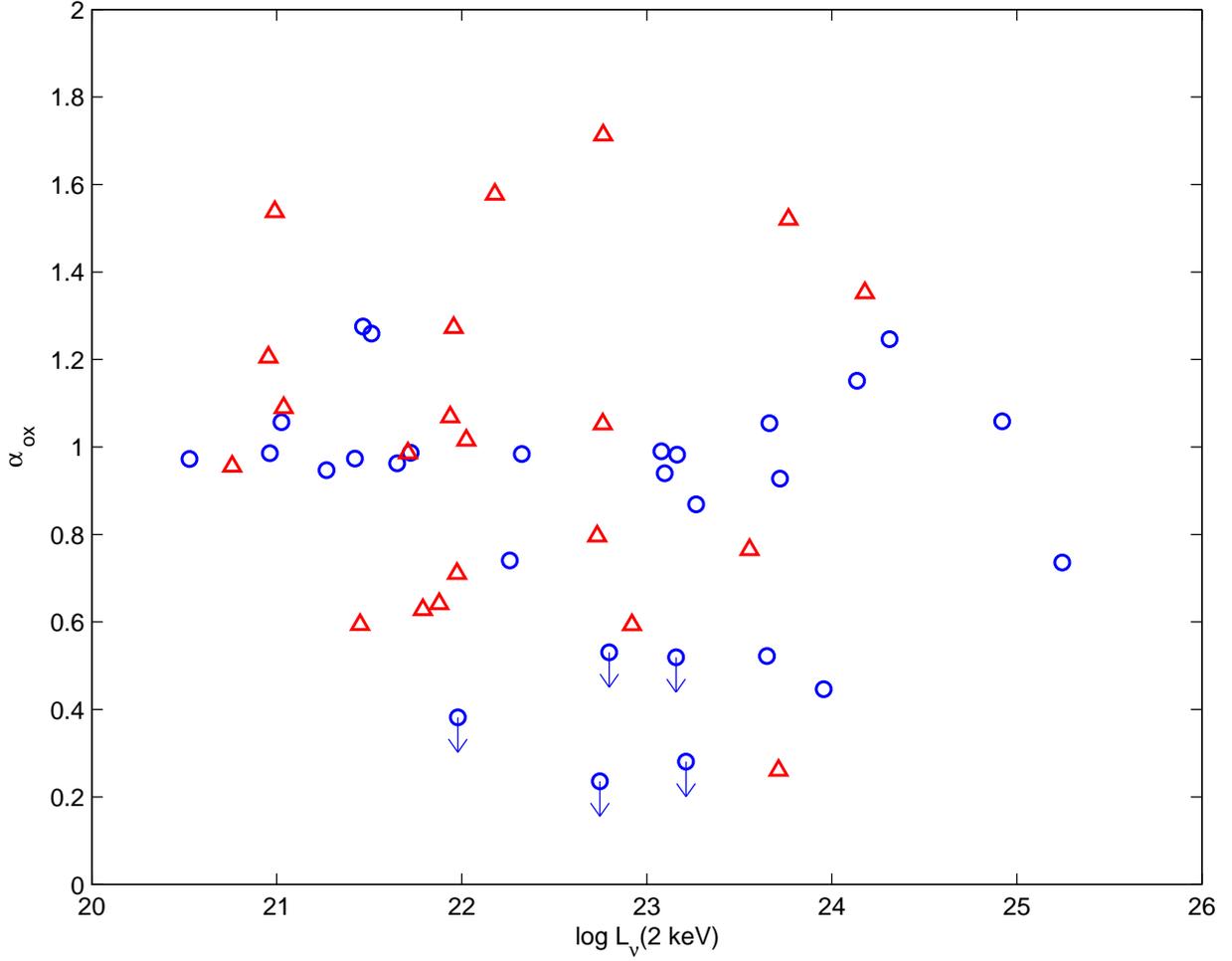} \caption{Optical/UV to X-ray spectral index
$\alpha_{\rm ox}$ vs. the X-ray spectral luminosity at 2 keV,
$L_{\nu({\rm 2~ keV})}$. The symbols are the same as Figure 1. }
\label{fig_alpha_lx}
\end{figure}

\begin{figure}
\plotone{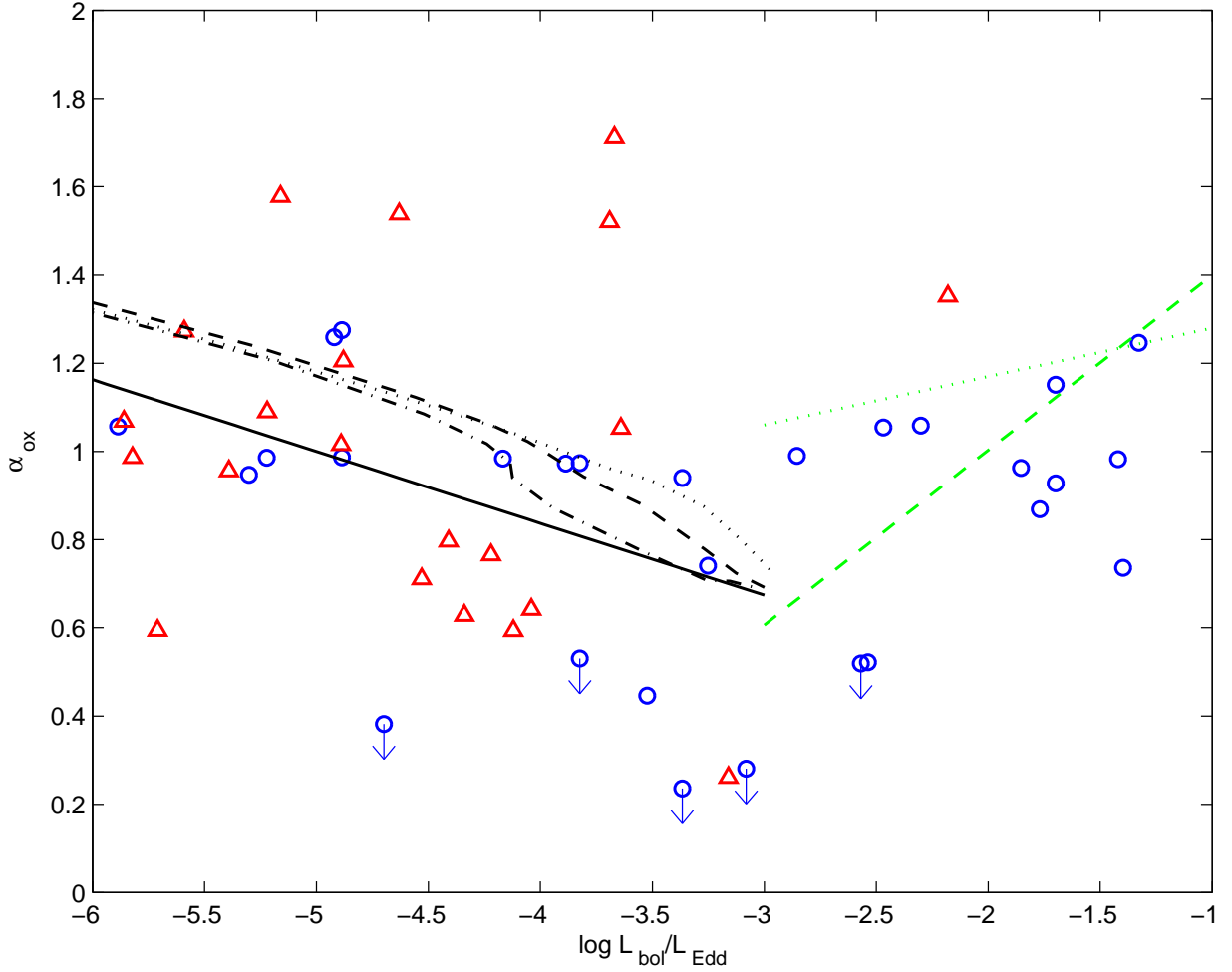} \caption{Optical/UV to X-ray spectral index
$\alpha_{\rm ox}$ vs. the Eddington ratio $L_{\rm bol}/L_{\rm Edd}$.
The symbols are the same as Figure 1. The strong anti-correlation
for the sources with $L_{\rm bol}/L_{\rm Edd}\lesssim 10^{-3}$ is
plotted in solid line. The green dashed and dotted lines correspond
to the correlations found by Lusso(2010) and Grupe(2010),
respectively. The black dotted, dashed, and dash-dotted lines
correspond to the simulated results assuming the black hole mass as
$10^7, 10^8$, and $10^9$$M_{\odot}$, respectively. }
\label{fig_alpha_redd}
\end{figure}

\begin{figure}
\plotone{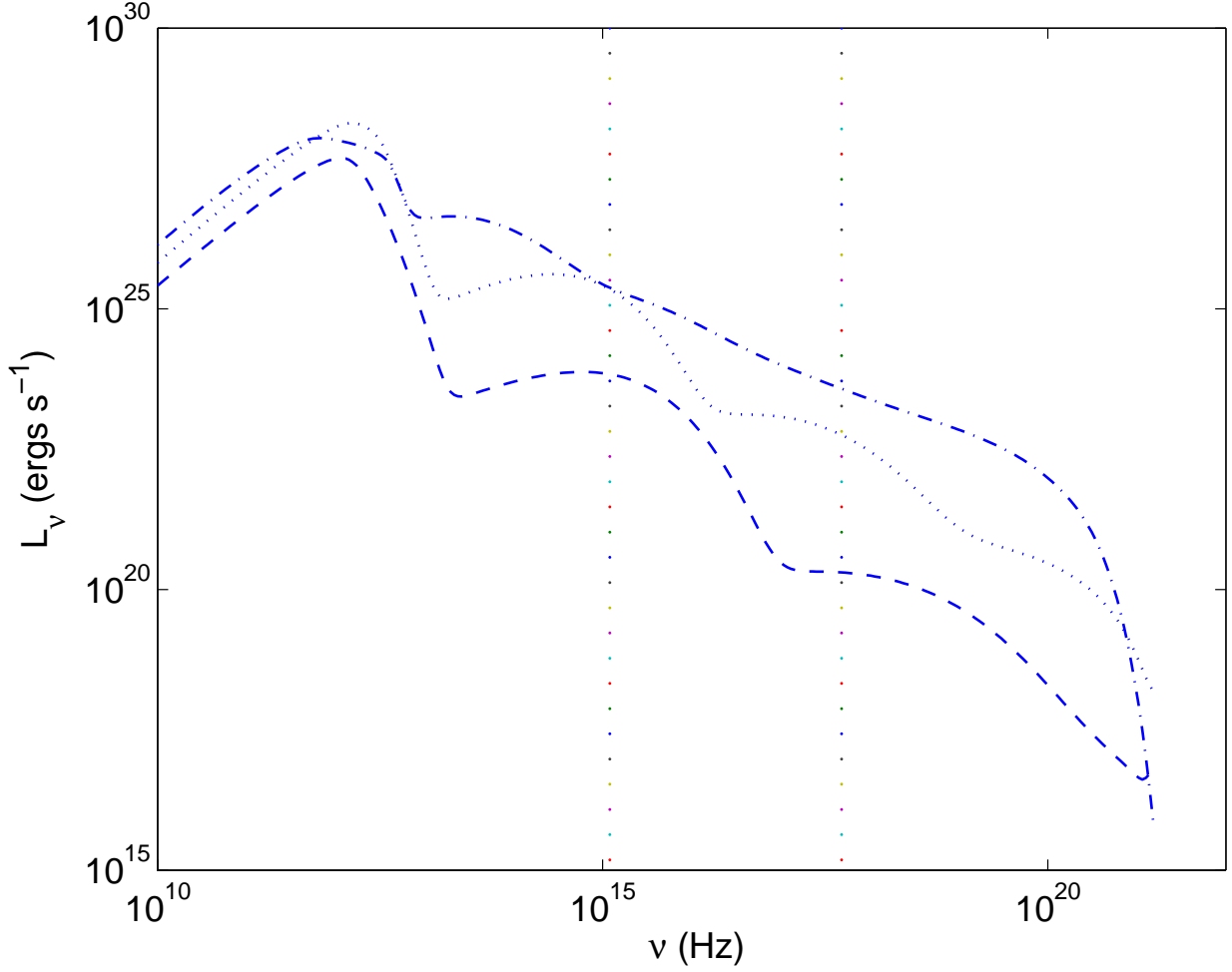} \caption{Spectrum emitted from the simulated
LLAGN with the ADAF model. The black hole mass is assumed to be
$10^8M_{\odot}$. The dashed, dotted, and dash-dotted lines represent
the spectra from the ADAF with the accretion rate $\dot{m}=10^{-4}$,
$10^{-3}$, and $10^{-2}$, respectively. The corresponding Eddington
ratios are $7.46\times 10^{-7}$, $1.06\times 10^{-4}$, and
$10^{-3}$, and the derived $\alpha_{\rm ox}$ for the three spectra
are 1.36, 1.00, and 0.69, respectively. The two vertical lines show
the frequencies at the wavelength $\lambda=2500~\rm \AA$ and energy
2 keV. } \label{fig_nu_lnu}
\end{figure}

\end{document}